# Quantum Information Transmission over a Partially Degradable Channel

Laszlo Gyongyosi, *Member, IEEE*

[1] Quantum Technologies Laboratory, Department of Telecommunications
*Budapest University of Technology and Economics*
2 Magyar tudosok krt, H-1111, Budapest, Hungary
[2] Information Systems Research Group, Mathematics and Natural Sciences
*Hungarian Academy of Sciences*
H-1518, Budapest, Hungary

gyongyosi@hit.bme.hu

**We investigate a quantum coding for quantum communication over a PD (partially degradable) degradable quantum channel. For a PD channel, the degraded environment state can be expressed from the channel output state up to a degrading map. PD channels can be restricted to the set of optical channels which allows for the parties to exploit the benefits in experimental quantum communications. We show that for a PD channel, the partial degradability property leads to higher quantum data rates in comparison to those of a degradable channel. The PD property is particular convenient for quantum communications and allows one to implement the experimental quantum protocols with higher performance. We define a coding scheme for PD-channels and give the achievable rates of quantum communication.**

The mathematical formalism that stands behind the information theoretic description of a quantum channel represent a flexible framework, which allows for one to study the quantum information conveying capabilities of noisy quantum links. In this work we reveal that more efficient quantum communication is possible over a PD (partially degradable [13-14]) channel in comparison to standard degradable quantum channels.

The definition of conjugate degradable quantum channels was introduced by Bradler *et al.* [1]. They showed that for a conjugate degradable channel the complex conjugated environment state $E'$ can be simulated from the channel output $B$. In their examples [1], the connection between environment states $E$ and $E'$ was a complex conjugation. For a PD channel the connection between $E$ and $E'$ is a degrading CPTP (Completely Positive Trace Preserving) map [13]. Currently, we have no result for the rates of quantum communication over a PD channel that can be achieved by polar coding. The polar coding [4] has been already studied in the quantum setting, and results have been obtained on the rates of classical and quantum communication over degradable and anti-degradable channels [6-8,12-13]. On the other hand, up to this date the possibilities of partial degradability are still unrevealed. Here we show that if a quantum channel is both degradable and has the PD property, then polar codes can produce better performance in comparison to degradable channels. As we have found, the rate of quantum communication with polar codes can be exceeded in comparison to the currently known results on degradable channels [7-8]. By exploiting the PD property, quantum communication will be possible in a regime nearer to the channel's quantum capacity [9-11] in comparison to polar codes that were developed for degradable channels.

This paper is organized as follows. First, we define the codeword sets for a PD channel, and then we give the rates of the quantum communication. In the Supplemental Material, we study the performance of the code.

For a PD channel, we use the following description (see Fig. 1.). The channel between Alice and Bob is denoted by $\mathcal{N}_{AB}$. The input and output of $\mathcal{N}_{AB}$ are denoted by $A$ and $B$. Since the channel is degradable, Bob can simulate the environment's channel $\mathcal{N}_{AE'}$ by applying the degrading map $\mathcal{D}^{B \to E'}$ on $B$, $\mathcal{N}_{AE'} = \mathcal{D}^{B \to E'} \circ \mathcal{N}_{AB}$. For the error-probabilities of a *degradable* channel $\mathcal{N}_{AB}$ (i.e., a *degraded* or *non-degradable* $\mathcal{N}_{AE'}$), the relation $p_{AE'} \geq p_{AB}$ holds [1,2-3,5]. Since the channel is PD, the output $E$ of the channel between Alice and the environment, called the complementary channel $\mathcal{N}_{AE}$, can be used to simulate the *degraded* environment state $E'$. Applying the *degrading* map $\mathcal{D}^{E \to E'}$ on $E$, the result is $E'$. From the PD property, an important corollary follows: the output $E$ of the complementary channel $\mathcal{N}_{AE}$ is not equivalent to the output $E'$ of channel $\mathcal{N}_{AE'}$, where $\mathcal{N}_{AE'} = \mathcal{N}_{AE} \circ \mathcal{D}^{E \to E'}$.

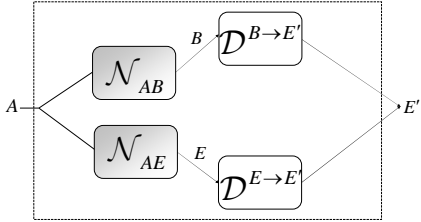

**FIG. 1.** A PD channel. The environment state $E$ is outputted by the complementary channel $\mathcal{N}_{AE}$. The degraded environment state will be referred to as $E'$, which is the output of channel $\mathcal{N}_{AE'} = \mathcal{N}_{AE} \circ \mathcal{D}^{E \to E'}$, where $\mathcal{D}^{E \to E'}$ is the degradation map. The environment state $E'$ contains less valuable information than state $E$, which makes it possible to use a PD channel with higher quantum communication rates in comparison to a degradable channel.

The polar coding scheme investigated for a PD channel $\mathcal{N}$, follows the basic polar coding scheme of [8] for degradable channels. The encoding of quantum information will use the amplitude and phase information. The amplitude "good" and "bad" codewords are depicted by $\mathcal{G}(\mathcal{N}_{amp}, \beta)$, $\mathcal{B}(\mathcal{N}_{amp}, \beta)$, and similarly for the phase encoding $\mathcal{G}(\mathcal{N}_{phase}, \beta)$, $\mathcal{B}(\mathcal{N}_{phase}, \beta)$. From these sets further polar codeword sets can be constructed. The density matrices of the frozen bits $\varsigma_{A_1}, \ldots, \varsigma_{A_{n-l}}$ belong to the set

$$\mathcal{G}(\mathcal{N}_{phase}, \beta) \cap \mathcal{B}(\mathcal{N}_{ampl}, \beta) \quad (1)$$

or

$$\mathcal{G}(\mathcal{N}_{ampl}, \beta) \cap \mathcal{B}(\mathcal{N}_{phase}, \beta). \quad (2)$$

The frozen bits from the set of

$$\mathcal{B}(\mathcal{N}_{ampl}, \beta) \cap \mathcal{B}(\mathcal{N}_{phase}, \beta) \quad (3)$$

require pre-shared entanglement. We extend the results on a degradable channel for a PD channel.

**Theorem 1**. *The frozen bits of the polar encoder $\mathcal{E}$ for a PD channel do not require pre-shared entanglement.*
*Proof.* The proof trivially follows from the previously derived results in [8]. Since our channel $\mathcal{N}$ is degraded, the rate of entanglement consumption between encoder $\mathcal{E}$ and Bob's decoder $\mathcal{D}$ is zero [8], i.e., $\lim_{n \to \infty} \frac{1}{n} |\mathcal{B}| = 0$.
∎

**Corollary 1.** (On the net rate for a PD channel). *Since for a PD channel the rate of entanglement consumption is zero, the net rate will be equal to the rate of quantum communication $R_Q(\mathcal{N})$.*

**Polar code sets for a PD channel.** To derive the $R_Q(\mathcal{N})$ rate of quantum communication for a PD channel $\mathcal{N}$, we define two sets that are "good" for a PD channel but "bad" for a degradable channel, $\mathcal{P}_1'$ and $\mathcal{P}_2'$ as follows:

$$\mathcal{P}_1' \subseteq \mathcal{P}_1, \ \mathcal{P}_2' \subseteq \mathcal{P}_2 \quad (4)$$

and the amplitude and phase frozen sets for a PD channel as

$$\mathcal{P}_1 \setminus \mathcal{P}_1', \ \mathcal{P}_2 \setminus \mathcal{P}_2', \quad (5)$$

where

$$\mathcal{P}_1 = \left(\mathcal{G}(\mathcal{N}_{amp}, \beta) \cap \mathcal{B}(\mathcal{N}_{phase}, \beta)\right)^{degr}, \quad (6)$$

$$\mathcal{P}_2 = \left(\mathcal{B}(\mathcal{N}_{amp}, \beta) \cap \mathcal{G}(\mathcal{N}_{phase}, \beta)\right)^{degr} \quad (7)$$

are the amplitude or phase frozen sets for a degradable channel, and $|S_{bad}| = n - (m + \Delta) = n - m - |\mathcal{P}_1'|$, where

$$\Delta = |\mathcal{P}_1| - |\mathcal{P}_1 \setminus \mathcal{P}_1'| = |\mathcal{P}_1'|, \quad (8)$$

along with

$$R_Q\left(\mathcal{N}_{AE}\right) - R_Q\left(\mathcal{N}_{AE'}\right) = \lim_{n\to\infty}\frac{1}{n}\left(\left|\mathcal{P}_1'\right|\right). \quad (9)$$

From the quantity of (8) trivially follows, that if $\mathcal{D}^{E\to E'}$ is equal to a complex conjugation $\mathcal{C}$, then $\Delta = 0$; for this case the performance of the code is equivalent to the codes that of degradable channels.

For a PD channel with degrading map $\mathcal{D}^{E\to E'}$ it results in a new "good", denoted by $S_{in}^{conj.PD}$, and defined as:

$$S_{in}^{conj.PD} = S_{in}^{degr} \cup \mathcal{P}_1' \\ = \left(\mathcal{G}\left(\mathcal{N}_{amp},\beta\right) \cap \mathcal{G}\left(\mathcal{N}_{phase},\beta\right)\right)^{conj.PD}, \quad (10)$$

with

$$\left|S_{in}^{conj.PD}\right| = \left|S_{in}^{degr}\right| + \Delta = m + \Delta, \quad (11)$$

where $m < m + \Delta$, and $\left|S_{in}^{conj.PD}\right| + \left|S_{bad}\right| = n = 2^k$, and $k$ is the level of the polar structure [3]. In comparison to a degradable channel, for a PD channel, the density matrices of the frozen bits $\varsigma_{A_1},\ldots,\varsigma_{A_{n-l}}$ will be selected from the sets $\mathcal{P}_1 \setminus \mathcal{P}_1'$ and $\mathcal{P}_2 \setminus \mathcal{P}_2'$, instead of $\mathcal{P}_1$ and $\mathcal{P}_2$. The bad set is defined as in the case of degradable channels $\mathcal{B} = \mathcal{B}\left(\mathcal{N}_{amp},\beta\right) \cap \mathcal{B}\left(\mathcal{N}_{phase},\beta\right)$. Set $\mathcal{B}$ will not be used as frozen bits, since the channel is degradable.

**Theorem 2.** (On the rates of polar coding for a PD channel). *For the non-empty set $\mathcal{P}_1'$, $\left|\mathcal{P}_1'\right| > 0$, the $R_Q$ rate of quantum communication for a PD channel with set $S_{in}^{conj.PD}$ is higher than the rate of quantum communication that can be obtained for a degradable channel with $S_{in}^{degr}$.*

*Proof.* The defined sets $\mathcal{P}_1'$ and $\mathcal{P}_2'$ for a PD channel are also disjoint [7-8,12], thus

$$\left|\mathcal{P}_1' \cup \mathcal{P}_2'\right| = \left|\mathcal{P}_1'\right| + \left|\mathcal{P}_2'\right|, \quad (12)$$

and since the elements of set $\mathcal{P}_2$ are not transmitted over the channel (Bob synthesizes $\mathcal{P}_2$ from the corresponding subset $\Omega_{\mathcal{P}_1}$ of $\mathcal{P}_1$ by using Hadamard operations in the decoding process):

$$\lim_{n\to\infty}\frac{1}{n}\left|\left(\mathcal{P}_2 \setminus \mathcal{P}_2'\right)\right| = 0 \quad (13)$$

and

$$\lim_{n\to\infty}\frac{1}{n}\left|\mathcal{P}_2'\right| = 0, \quad (14)$$

along with relations $\left|\left(\mathcal{P}_2 \setminus \mathcal{P}_2'\right) \cap \mathcal{G}\left(\mathcal{N}_{amp},\beta\right)\right| = 0$, $\left|\mathcal{P}_2' \cap \mathcal{G}\left(\mathcal{N}_{amp},\beta\right)\right| = 0, \left|\mathcal{B}\left(\mathcal{N}_{amp},\beta\right) \cap \mathcal{G}\left(\mathcal{N}_{amp},\beta\right)\right| = 0$, which follows from the fact that $\mathcal{P}_2' \subseteq \mathcal{P}_2$ and

$$\left(\mathcal{P}_2 \setminus \mathcal{P}_2'\right) \subseteq \mathcal{B}\left(\mathcal{N}_{amp},\beta\right), \quad (15)$$

where

$$\mathcal{B}\left(\mathcal{N}_{amp},\beta\right) = [n] \setminus \mathcal{G}\left(\mathcal{N}_{amp},\beta\right) \quad (16)$$

and

$$\left|\left(\mathcal{P}_2 \setminus \mathcal{P}_2'\right) \cap \left(S_{in}^{conj.PD} \cup \mathcal{B}\right)\right| = 0, \quad (17)$$

$$\mathcal{P}_1' \cup \mathcal{P}_2' \subseteq \mathcal{G}\left(\mathcal{N}_{amp},\beta\right), \quad (18)$$

and

$$\left(\mathcal{P}_1 \setminus \mathcal{P}_1'\right) \cup \left(\mathcal{P}_2 \setminus \mathcal{P}_2'\right) \subseteq \mathcal{G}\left(\mathcal{N}_{amp},\beta\right), \quad (19)$$

see (13) and (14). For a PD channel $\mathcal{P}_1' \subseteq \mathcal{G}\left(\mathcal{N}_{amp},\beta\right)$ and $\left(\mathcal{P}_1 \setminus \mathcal{P}_1'\right) \subseteq \mathcal{G}\left(\mathcal{N}_{amp},\beta\right)$, which demonstrates that the defined codewords sets $\mathcal{P}_1'$, $\mathcal{P}_2'$, $\mathcal{P}_1 \setminus \mathcal{P}_1'$ and $\mathcal{P}_2 \setminus \mathcal{P}_2'$ are pairwise disjoint, where

$$\lim_{n\to\infty}\frac{1}{n}\left|\mathcal{G}\left(\mathcal{N}_{amp},\beta\right)\right| + \\ \lim_{n\to\infty}\frac{1}{n}\left|[n] \setminus \left(\left(\mathcal{P}_1' \cup \mathcal{P}_2'\right) \cup \left(\mathcal{P}_1 \setminus \mathcal{P}_1'\right) \cup \left(\mathcal{P}_2 \setminus \mathcal{P}_2'\right)\right)\right| = 1, \quad (20)$$

and

$$\left|\mathcal{G}\left(\mathcal{N}_{amp},\beta\right) \setminus \mathcal{G}\left(\mathcal{N}_{amp},\beta\right) \cap \mathcal{G}\left(\mathcal{N}_{phs},\beta\right)\right| + \left|\left(\mathcal{P}_2 \setminus \mathcal{P}_2'\right)\right| \\ + \left|[n] \setminus \left(\left(\mathcal{P}_1' \cup \mathcal{P}_2'\right) \cup \left(\mathcal{P}_1 \setminus \mathcal{P}_1'\right) \cup \left(\mathcal{P}_2 \setminus \mathcal{P}_2'\right)\right)\right| \leq n \quad (21)$$

with

$$\left([n] \setminus \left(\mathcal{P}_1'\right)\right) \subseteq \left(\mathcal{G}\left(\mathcal{N}_{amp},\beta\right) \cap \mathcal{G}\left(\mathcal{N}_{phase},\beta\right)\right)^{conj.PD} \\ \cup \left(\left(\mathcal{P}_1 \setminus \mathcal{P}_1'\right)\right). \quad (22)$$

Using the codeword construction defined for the PD channel, the $R_Q\left(\mathcal{N}\right)$ rate can be achieved over the channel $\mathcal{N}_{AB}$ can be expressed as:

$$R_Q\left(\mathcal{N}\right) = \lim_{n\to\infty}\frac{1}{n}\left(\left|S_{in}^{degr} \cup \mathcal{P}_1' \cup \mathcal{P}_2'\right|\right) \\ = \lim_{n\to\infty}\frac{1}{n}\left|\left(\mathcal{G}\left(\mathcal{N}_{ampl},\beta\right) \cap \mathcal{G}\left(\mathcal{N}_{phs},\beta\right)\right)^{degr} \cup \mathcal{P}_1' \cup \mathcal{P}_2'\right|.$$

Assuming $\beta < 0.5$, the following relation holds for the Holevo information of channel $\mathcal{N}_{AB}$ and $\mathcal{N}_{AE'}$ of $\mathcal{N}$:

$$\chi\left(\mathcal{N}_{AB}\right) = \lim_{n\to\infty}\frac{1}{n}\left|\mathcal{G}\left(\mathcal{N}_{amp},\beta\right)\cup\mathcal{P}_2'\right|, \qquad (23)$$

$$\chi\left(\mathcal{N}_{AE}\right) = \lim_{n\to\infty}\frac{1}{n}\left(\left|\mathcal{P}_1\right|+\left|\mathcal{P}_2\right|\right) \qquad (24)$$

$$\chi\left(\mathcal{N}_{AE'}\right) = \lim_{n\to\infty}\frac{1}{n}\left(\left|\left(\mathcal{P}_1\setminus\mathcal{P}_1'\right)\right|+\left|\left(\mathcal{P}_2\setminus\mathcal{P}_2'\right)\right|\right), \qquad (25)$$

where $\chi\left(\mathcal{N}_{AB}\right)$, $\chi\left(\mathcal{N}_{AE}\right)$ and $\chi\left(\mathcal{N}_{AE'}\right)$ are the Holevo information of the channels $\mathcal{N}_{AB}$, $\mathcal{N}_{AE}$ and $\mathcal{N}_{AE'}$ of $\mathcal{N}$. Combing this result with (13) and (14), one will get

$$\chi\left(\mathcal{N}_{AE'}\right) = \lim_{n\to\infty}\frac{1}{n}\left(\left|\left(\mathcal{P}_1\setminus\mathcal{P}_1'\right)\right|\right) \qquad (26)$$

and the rate of quantum communication is (*Note: no maximization and regularization needed in the first line of* (27), *since in terms of polar coding the quantum capacity is symmetric and it is additive for any degradable channel* [1,5-6,8])

$$\begin{aligned} R_Q\left(\mathcal{N}\right) &= \chi\left(\mathcal{N}_{AB}\right)-\chi\left(\mathcal{N}_{AE'}\right) \\ &= \lim_{n\to\infty}\frac{1}{n}\left(\left|\mathcal{G}\left(\mathcal{N}_{amp},\beta\right)\right|-\left|\left(\mathcal{P}_1\setminus\mathcal{P}_1'\right)\right|\right). \end{aligned} \qquad (27)$$

The result obtained in (27) can be rewritten as follows:

$$R_Q\left(\mathcal{N}\right) = \lim_{n\to\infty}\frac{1}{n}\left(\left|S_{in}^{degr}\right|+\left|\mathcal{P}_1'\right|\right) = \lim_{n\to\infty}\frac{1}{n}\left(\left|S_{in}^{conj.PD}\right|\right). \qquad (28)$$

From the polar encoding scheme, it follows that for $\beta < 0.5$:

$$\sqrt{F\left(S_{in}^{conj.PD}\cup\left(\mathcal{P}_2\setminus\mathcal{P}_2'\right)\right)} < 2^{-n^\beta}, \qquad (29)$$

and for the fidelity parameters of $\mathcal{P}_1$:

$$\sqrt{F\left(\left(\mathcal{P}_1\setminus\mathcal{P}_1'\right)\right)} \geq 1-2^{-n^\beta}. \qquad (30)$$

If (29) and (30) hold, then $\left(\mathcal{P}_1\setminus\mathcal{P}_1'\right)\cap\mathcal{B} \neq \varnothing$. Using the defined sets it follows that

$$\begin{aligned} &\left(\mathcal{B}\left(\mathcal{N}_{amp},\beta\right)\cap\mathcal{B}\left(\mathcal{N}_{phase},\beta\right)\right) \subseteq \\ &\left(S_{in}^{conj.PD}\cup\left(\mathcal{B}\left(\mathcal{N}_{amp},\beta\right)\cap\mathcal{B}\left(\mathcal{N}_{phase},\beta\right)\right)\right) = \varnothing. \end{aligned} \qquad (31)$$

After some steps of reordering, we get that

$$\left(\left(\mathcal{P}_1\setminus\mathcal{P}_1'\right)\right)\cap\left(\mathcal{B}\left(\mathcal{N}_{amp},\beta\right)\cap\mathcal{B}\left(\mathcal{N}_{phase},\beta\right)\right) \subseteq \\ \left(\left(\mathcal{P}_1\setminus\mathcal{P}_1'\right)\right)\cap\left(S_{in}^{conj.PD}\cup\begin{pmatrix}\mathcal{B}\left(\mathcal{N}_{amp},\beta\right)\\ \cap\mathcal{B}\left(\mathcal{N}_{phase},\beta\right)\end{pmatrix}\right) = \varnothing.$$

and

$$\left(\mathcal{P}_1\setminus\mathcal{P}_1'\right)\cap\left(S_{in}^{conj.PD}\cup\left(\mathcal{P}_2\setminus\mathcal{P}_2'\right)\right) = \varnothing. \qquad (32)$$

Since channel $\mathcal{N}_{AE'}$ is degraded [8],

$$\lim_{n\to\infty}\frac{1}{n}\left|\mathcal{B}\right| = 0. \qquad (33)$$

These results also mean that the codeword sets $S_{in}^{degr}$, $S_{in}^{conj.PD}$, $\mathcal{P}_1' \subseteq \mathcal{P}_1$, $\mathcal{P}_2' \subseteq \mathcal{P}_2$ and $\mathcal{B}$ are disjoint sets with relation

$$\begin{aligned} &\left|S_{in}^{degr}\cup\mathcal{P}_1'\cup\mathcal{P}_2'\cup\left(\mathcal{P}_1\setminus\mathcal{P}_1'\right)\cup\left(\mathcal{P}_2\setminus\mathcal{P}_2'\right)\right| \\ &= \left|S_{in}^{conj.PD}\cup\left(\mathcal{P}_1\setminus\mathcal{P}_1'\right)\cup\left(\mathcal{P}_2\setminus\mathcal{P}_2'\right)\right| \\ &= \left|S_{in}^{conj.PD}\cup\left(\mathcal{P}_1\setminus\mathcal{P}_1'\right)\right|, \end{aligned} \qquad (34)$$

hence for a PD channel $\mathcal{N}$ we get:

$$R_Q\left(\mathcal{N}\right) = \lim_{n\to\infty}\frac{1}{n}\left(\left|S_{in}^{degr}\cup\mathcal{P}_1'\right|\right) = \lim_{n\to\infty}\frac{1}{n}\left(\left|S_{in}^{conj.PD}\right|\right).$$

It can be rewritten as follows:

$$\begin{aligned} R_Q\left(\mathcal{N}\right) &= \lim_{n\to\infty}\frac{1}{n}\left(\left|S_{in}^{degr}\right|+\left|\mathcal{P}_1'\right|\right) = \\ &= \lim_{n\to\infty}\frac{1}{n}\left(\left|\mathcal{G}\left(\mathcal{N}_{amp},\beta\right)\right|-\left|\left(\mathcal{P}_1\setminus\mathcal{P}_1'\right)\right|\right) \\ &= \lim_{n\to\infty}\frac{1}{n}\left(\left|S_{in}^{conj.PD}\right|\right). \end{aligned} \qquad (35)$$

The available codewords for quantum communication over a PD channel $\mathcal{N}_{AB}$ will be

$$\left|S_{in}^{conj.PD}\right| = \left|S_{in}^{degr}\right|+\left|\mathcal{P}_1'\right|, \qquad (36)$$

which concludes our proof. These results show that for the non-empty set $\mathcal{P}_1'$, the results on rate of quantum communication $R_Q\left(\mathcal{N}\right)$ for a PD channel $\mathcal{N}$ exceed the result obtained for a degradable channel with $\left|S_{in}^{degr}\right|$.

∎

**Summary.** Since for a degradable channel $\mathcal{N}$, only set $S_{in}^{degr} = \left(\mathcal{G}\left(\mathcal{N}_{ampl},\beta\right)\cap\mathcal{G}\left(\mathcal{N}_{phs},\beta\right)\right)^{degr}$ can be used for quantum communication, it will result in quantum data rate

$$\begin{aligned} R_Q\left(\mathcal{N}\right) &= \lim_{n\to\infty}\frac{1}{n}\left(\left|S_{in}^{degr}\right|\right) \\ &= \lim_{n\to\infty}\frac{1}{n}\left|\left(\mathcal{G}\left(\mathcal{N}_{amp},\beta\right)\cap\mathcal{G}\left(\mathcal{N}_{phase},\beta\right)\right)^{degr}\right|. \end{aligned} \qquad (37)$$

The codewords that can transmit amplitude information are depicted by $\mathcal{G}\left(\mathcal{N}_{amp},\beta\right)$. The set that can transmit phase is denoted by $\mathcal{G}\left(\mathcal{N}_{phs},\beta\right)$. For a degradable channel $\mathcal{N}$ only the set $\left|S_{in}^{degr}\right| = m$ can be used for quan-tum communication. For a degradable channel $\mathcal{N}$, valuable information can be leaked only from the set $\mathcal{P}_1$ [7-8] which results in $\left( S_{in}^{degr} \cup \mathcal{P}_1 \cup \mathcal{P}_2 \right) \setminus \left( \mathcal{P}_1 \cup \mathcal{P}_2 \right)$, where $\lim_{n \to \infty} \frac{1}{n} |\mathcal{P}_2| = 0$, i.e., $\left( S_{in}^{degr} \cup \mathcal{P}_1 \right) \setminus \left( \mathcal{P}_1 \right)$. The set $\mathcal{P}_1$ represents the information of channel $\mathcal{N}_{AE}$, between Alice and the environment. For these sets $R_Q(\mathcal{N}_{AE}) > 0$ and $R_Q(\mathcal{N}_{AB}) = 0$. As we have found, this is not the case, if the degradable channel is also partially degradable. While for a degradable channel, only the set $S_{in}^{degr}$ can be used for quantum communication, the situation will change for a PD channel, since this property causes changes in the information of complementary channel. In this case, the achievable codeword set for quantum communication is

$$\begin{aligned} S_{in}^{conj.PD} &= S_{in}^{degr} \cup \mathcal{P}_1' \cup \mathcal{P}_2' \\ &= \left( \mathcal{G}(\mathcal{N}_{ampl}, \beta) \cap \mathcal{G}(\mathcal{N}_{phs}, \beta) \right)^{degr} \cup \mathcal{P}_1' \cup \mathcal{P}_2' \\ &= \left( \mathcal{G}(\mathcal{N}_{ampl}, \beta) \cap \mathcal{G}(\mathcal{N}_{phs}, \beta) \right)^{conj.PD}, \end{aligned} \tag{38}$$

where $\mathcal{P}_1' \subseteq \mathcal{P}_1$, $\mathcal{P}_2' \subseteq \mathcal{P}_2$ can be used for quantum communication, which results in quantum data rate

$$\begin{aligned} R_Q(\mathcal{N}) &= \lim_{n \to \infty} \frac{1}{n} \left( \left| S_{in}^{degr} \cup \mathcal{P}_1' \right| \right) \\ &= \lim_{n \to \infty} \frac{1}{n} \left( \left| S_{in}^{conj.PD} \right| \right). \end{aligned} \tag{39}$$

For a PD channel the extended set $S_{in}^{conj.PD} = S_{in}^{degr} \cup \mathcal{P}_1' \cup \mathcal{P}_2'$ with $\left| S_{in}^{conj.PD} \right| = \left| S_{in}^{degr} \right| + \Delta = m + \Delta$ will be available for quantum communication, where $\mathcal{P}_1' \subseteq \mathcal{P}_1$, $\mathcal{P}_2' \subseteq \mathcal{P}_2$, and $\lim_{n \to \infty} \frac{1}{n} |\mathcal{P}_2 \setminus \mathcal{P}_2'| = 0$ and $\lim_{n \to \infty} \frac{1}{n} |\mathcal{P}_2'| = 0$.

The extended set results in higher quantum communication rates in comparison to a degradable one. The amount of information in the degraded environment state $E'$ outputted by channel $\mathcal{N}_{AE'} = \mathcal{N}_{AE} \circ \mathcal{D}^{E \to E'}$ is less than in the environment state $E$, outputted by channel $\mathcal{N}_{AE}$. This property results in better quantum communication rates over PD quantum channels in comparison to a quantum channels that has no the PD property. The results allow to implement quantum protocols over quantum links with higher performance than it is currently available by the standard quantum codes.

**Acknowledgements**

The results discussed above are supported by the grant TAMOP-4.2.2.*B*-10/1--2010-0009 and COST Action MP1006.

# Supplemental Information

**Performance of the Coding Scheme.** To study the performance of the polar coding scheme over a PD channel, we will use the $1 \to N$ cloning channel [1,8], where $\mathcal{D}^{E \to E'}$ is equal to a complex conjugation $\mathcal{C}$, and $\Delta = 0$. The $p_{BER}$ values are calculated by the $F_i$ fidelity parameters of the resulting $2^k$ logical $1 \to N$ cloning channels $\sum_{i=1}^{2^k} \mathcal{N}_i$ in the polarized $k$-level channel structure, assuming input block code length $n = 2^k$. The $F(\mathcal{N}_i)$ fidelity of the $i$-th logical $1 \to N$ cloning channel $\mathcal{N}_i$ is $F(\mathcal{N}_i) \to 0$, if it belongs to the "good" logical channel set and $F(\mathcal{N}_i) \to 1$ if it belongs to the "bad" (*"bad" set*: cannot transmit amplitude and phase simultaneously) logical channel set [3,7-8,12]. The *lower bound* of $p_{BER}$ is given as $p_{BER} \geq \frac{1}{2}\left(1 - \sqrt{1 - \sum_{i=1}^{2^k} F(\mathcal{N}_i)}\right)$, where $F(\mathcal{N}_i)$ is the fidelity of the $i$-th logical $1 \to N$ cloning channel $\mathcal{N}_i$. Using a threshold parameter $0 < \eta < 1$ on the $F(\mathcal{N}_i)$ fidelities of the $2^k$ logical $1 \to N$ cloning channels, the set of analyzed channels can be restricted to the set $\mathcal{A}(\eta) \equiv \{i : F(\mathcal{N}_i) \leq \eta\}$. Set $\mathcal{A}(\eta)$ will contain only those $1 \to N$ cloning channels for which $F(\mathcal{N}_i) \leq \eta$. Using the restricted set $\mathcal{A}(\eta)$ of channels, the *upper bound* on $p_{BER}$ can be evaluated as $p_{BER} \leq \frac{1}{2}\left(1 - \sqrt{\sum_{\mathcal{A}(\eta)} F(\mathcal{N}_i)}\right)$, where $F(\mathcal{N}_i)$ is the fidelity of the $i$-th $1 \to N$ cloning channel $\mathcal{N}_i$ from set $\mathcal{A}(\eta)$. An equality on the $p_{BER}$ block error probability also follows from Arikan's results [3], the encoded state $\rho_{u_i}$ in the given sequence $\rho_{u_1}^{n^k-1} \rho_{u_{n^k}}$, by using projectors $\Lambda_{(i)}^{F^{2^k}} : \Lambda_{(1),u_1}^{F^{2^k}}, \ldots, \Lambda_{(2^k),u_1^{2^k-1} u_{2^k}}^{F^{2^k}}$, the equality follows for the block error probability $p_{BER}$:

$$p_{BER} = \sum_{u_1^{i-1}} \frac{1}{2^{i-1}} \sum_{u_i} \frac{1}{2} Tr\left(\left(I - \Lambda_{(i),u_1^{i-1} u_i}^{F^{2^k}}\right) \sum_{u_{i+1}^{2^k}} \frac{1}{2^{2^k-1}} \rho_{u^{2^k}}\right),$$

where $I$ is the identity operator.

The upper bounds on block error probabilities for different values of $k$ for a cloning channel $1 \to N$ for $N = 1, 2, 3, 5, 8, 12$ and $24$, assuming maximally entangled inputs, are depicted in Fig. S.1. Assuming low BERs, $p_{BER} \leq 10^{-4}$ for a practical communication system, the rate of quantum communication $R_Q(\mathcal{N})$ of a $1 \to N$ cloning channel $\mathcal{N}$ can be nearly reached for moderate encoding complexity $\mathcal{O}(n \log n)$ with $n = 2^{10}$, and $k = 10$.

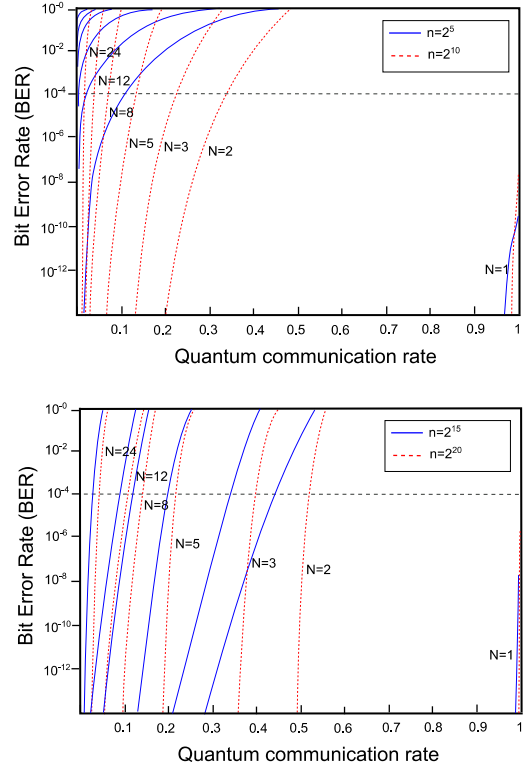

**FIG. S.1.** Block error probabilities and rate of quantum communication of a $1 \to N$ cloning channel for $N = 1, 2, 3, 5, 8, 12$ and $24$, assuming maximally entangled input. The number of channel uses were chosen to $n = 2^5, 2^{10}, 2^{15}$ and $2^{20}$. The $x$-axis represents the $R_Q(\mathcal{N})$ rate of quantum communication measured in qubits per channel use. The $y$-axis depicts the block error probabilities. Next to each dashed curve, we also show the $N$ to which the corresponding block error probability holds. Assuming low BER with $p_{BER} \leq 10^{-4}$ and $k = 10$, high performance code can be generated with low time complexity $\mathcal{O}(2^{10} \log 2^{10})$.

At $k \geq 20$, the rate $R_Q(\mathcal{N})$ will be near to its quantum capacity, i.e., $R_Q(\mathcal{N}) \approx Q(\mathcal{N})$. Below this level of $k$, only lower performances can be achieved by the polar encoding scheme, i.e., $R_Q(\mathcal{N}) < Q(\mathcal{N})$, which makes no possible to use an $1 \to N$ cloning channel for quantum communication in a regime near to its quantum capacity $Q(\mathcal{N})$. For moderate $k$ values, the encoding can be done in time $\mathcal{O}(n \log n)$. As we have found, for low encoding time complexity $\mathcal{O}(n \log n)$ with $n = 2^{10}$, and $k$=10, high rates can be achieved with $p_{BER} \leq 10^{-4}$, which makes possible to use the code with high efficiency over a PD channel.